\documentclass[sigconf]{acmart}
\usepackage{subcaption}

\copyrightyear{2026}
\acmYear{2026}
\setcopyright{cc}
\setcctype{by}
\acmConference[PACT '26]{International Conference on Parallel Architectures and Compilation Techniques}{October 19--22, 2026}{Chicago, IL, USA}
\acmBooktitle{International Conference on Parallel Architectures and Compilation Techniques (PACT '26), October 19--22, 2026, Chicago, IL, USA}
\acmDOI{10.1145/3838684.3847309}
\acmISBN{979-8-4007-2915-7/2026/10}

\begin{document}

%contenders - Chip of the old block, Chockablock, blockparty, blockwork
%Prototyping the Quantization-Perplexity Pareto Frontier on GPGPU Tensor Cores
%Pushing the Quantization-Perplexity Pareto Frontier with Custom Microscaling Format Exploration on Tensor Cores
%Pushing the Limits of Quantization with Custom Microscaling Format Exploration on GPGPU Tensor Cores
%Enabling Custom Microscaling Format Exploration on GPGPU Tensor Cores
\title{RoadBlock: Rethinking GPU Tensor Core Microarchitecture for Emerging Microscaling Format Support}

\author{Nikhil Rout}
\affiliation{%
  \institution{University of California, Los Angeles}
  \city{Los Angeles}
  \state{CA}
  \country{USA}}
\email{nikhilrout97@ucla.edu}

\author{Blaise Tine}
\affiliation{%
  \institution{University of California, Los Angeles}
  \city{Los Angeles}
  \state{CA}
  \country{USA}
  }
\email{blaisetine@cs.ucla.edu}

\begin{abstract}
While multiple recent custom Microscaling (MX) format proposals preserve higher model quality over OCP MX, their hardware evaluations remain confined to standalone MAC-unit prototypes, ignoring the system-level integration overheads inside real GPU Tensor Cores. To this end, we present RoadBlock, an open-source framework for accurate end-to-end modeling of custom MX formats in a programmable Tensor Core, built on the RISC-V Vortex GPGPU. RoadBlock exposes data size, scale-factor size, and block size as configurable parameters and automatically derives per-thread metadata tiling for arbitrary formats, enabling format-specific Perplexity-PPA tradeoff evaluation before silicon commitment.
\end{abstract}

\ccsdesc[500]{Computer systems organization~Parallel architectures}

\keywords{GPU Microarchitecture, Tensor Core, Quantization, Microscaling}

\maketitle

\section{Introduction}
Emerging Microscaling (MX) formats such as 4over6 \cite{cook2026sixaccuratenvfp4quantization}, IF4 \cite{cook2026adaptiveblockscaleddatatypes}, and RaZeR \cite{chen2026razerpushinglimitsnvfp4} are typically evaluated as standalone MAC-unit prototypes intended for systolic-array accelerators, silently ignoring the system-level integration overheads in NVIDIA Blackwell and AMD CDNA4-like GPU Tensor Cores that production ML systems stacks actually use. To bridge this gap, we propose \textbf{\textit{RoadBlock}}\footnote[1]{\href{https://github.com/vortexgpgpu/vortex/tree/roadblock}{\color{blue}https://github.com/vortexgpgpu/vortex/tree/roadblock}}, an open-source framework for end-to-end custom MX format modeling and Perplexity-PPA trade-off analysis, built on the programmable RISC-V Vortex GPGPU \cite{10.1145/3466752.3480128}. It exposes data size, scale-factor size, and block size as configurable parameters for MX format design-space exploration, and automatically derives per-thread scale-factor metadata tiling and register requirements for arbitrary formats.

\section{RoadBlock Microarchitecture}
%Fig.~\ref{fig:roadblock_overview} shows the extensions RoadBlock makes to a Vortex GPGPU SIMT sub-core to support custom MX formats in its Tensor Cores.

\begin{figure}
    \centering
    \includegraphics[width=0.9\linewidth]{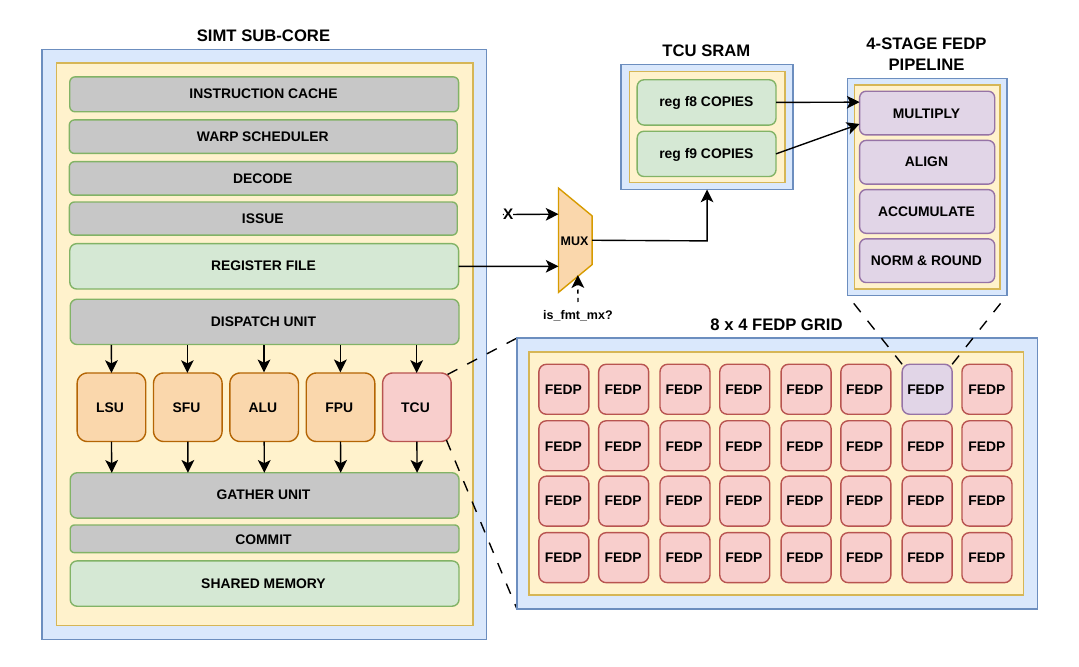}
    \caption{RoadBlock-extended Vortex GPGPU SIMT Sub-Core}
    \Description{RoadBlock-extended Vortex GPGPU SIMT Sub-Core}
    \label{fig:roadblock_overview}
\end{figure}

\paragraph{\textbf{MX-enhanced Fused Dot Product (FEDP)}}
The dot product for two MX-compliant vectors $A$ and $B$ of length $k$ is given by:
\begin{equation}
    \text{Dot}(A, B) = X^{(A)} X^{(B)} \sum_{i=1}^{k} \left( P_i^{(A)} \times P_i^{(B)} \right)
\end{equation}
where $X^{(A)}$, $X^{(B)}$ are block scale factors and $P_i^{(A)}$, $P_i^{(B)}$ are their corresponding elements. RoadBlock adopts and enhances the 4-cycle Ten-Four \cite{rout2026tenfouropensourcefuseddot} mixed-precision FEDP microarchitecture as illustrated in Fig.~\ref{fig:roadblock_overview}. Since Ten-Four performs early accumulation of the ``C'' addend alongside the vector products, the required scale-factor computations and shifts need to be applied to each lane's significand multiplication result in the first \texttt{MULTIPLY} stage itself.

\begin{figure*}[h]
  \centering
  \begin{subfigure}[t]{0.49\textwidth}
    \centering
    \includegraphics[width=\textwidth]{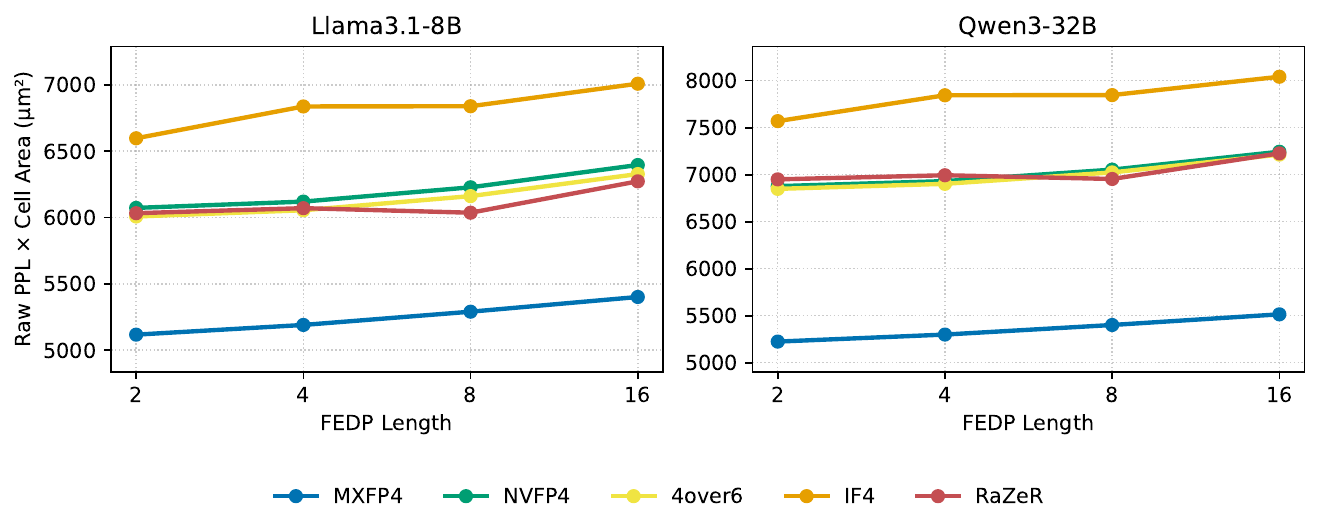}
    \caption{Perplexity $\times$ Cell Area ($\downarrow$)}
    \label{fig:roadblock-ppl-area}
  \end{subfigure}
  \hfill
  \begin{subfigure}[t]{0.49\textwidth}
    \centering
    \includegraphics[width=\textwidth]{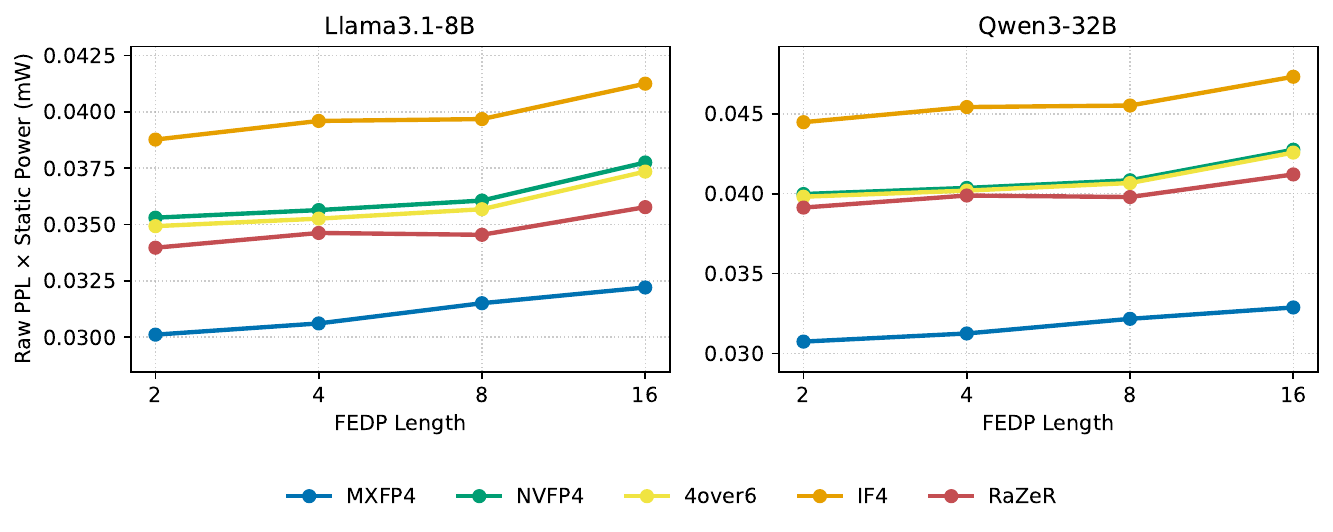}
    \caption{Perplexity $\times$ Static Power ($\downarrow$)}
    \label{fig:roadblock-ppl-static-power}
  \end{subfigure}
  \caption{WikiText-2 Perplexity -- Hardware Efficiency Tradeoff in Emerging 4-bit MX Formats Modeled on RoadBlock}
  \label{fig:roadblock-area-power-tradeoff}
\end{figure*}

\paragraph{\textbf{MX Scale Factor Metadata Tiling}}
Vortex's RISC-V-extended SIMT programming model provides a 32 registers/thread budget that is divided between matrices A, B, and C/D (8 each). Given that contiguous low-precision elements are packed along \texttt{K} into 32-bit bundles, Fig.~\ref{fig:mx_metadata_derivation} derives the required per-thread MX metadata for an arbitrarily designed format. Assuming NVFP4 (\texttt{d=4,w=8,k=16}) as the finest granularity, these formulae indicate a maximum requirement of 64 bits/thread (two extra registers) for MX metadata.

\begin{figure}
\footnotesize
\hrule
\vspace{3pt}
{\centering\textbf{\textsc{Given}}\par}

{\raggedright
$\mathit{No.\ Regs/Operand\ (NR)=8},\;
\mathit{No.\ Threads/Warp\ (NT)\in\{4,8,16,32\}}$

$\;d=\textit{element bits},\;
w=\textit{scale bits},\;
k=\textit{block size}$\par}
\vspace{3pt}

\hrule
\vspace{3pt}
\noindent\textbf{\textsc{Tiling (per-warp work chunk)}}
\begin{flalign}
&\mathit{tile\_cap}
= \mathit{NT}\times\mathit{NR}
= 8\,\mathit{NT} \quad
\ell_g
= \log_2(\mathit{tile\_cap})
= 3+\log_2(\mathit{NT}) &\\
&\mathit{tileM}
= 2^{\,\ell_g-\lfloor\ell_g/2\rfloor}
= \sqrt{8\,\mathit{NT}} \quad
\mathit{tileK}
= \tfrac{\mathit{tile\_cap}}
{\max(\mathit{tileM},\,\mathit{tileN})}
= \sqrt{8\,\mathit{NT}} &
\end{flalign}

\hrule
\vspace{3pt}
\noindent\textbf{\textsc{MX block scaling}}
\begin{flalign}
&\mathit{tileK}_{\mathit{fmt}}
= \mathit{tileK}\!\times\!\tfrac{32}{d}
= \tfrac{32\sqrt{8\,\mathit{NT}}}{d} \quad \mathit{num\_blocks}_{k}
= \frac{\mathit{tileK}_{\mathit{fmt}}}{k}
= \tfrac{32\sqrt{8\,\mathit{NT}}}{d\,k} &\\
&\mathit{scale\_bits}_{k}
= \mathit{num\_blocks}_{k}\times w
= \tfrac{32\,w\,\sqrt{8\,\mathit{NT}}}{d\,k} &
\end{flalign}

\hrule
\vspace{3pt}
\noindent\textbf{\textsc{MX Metadata}}
\begin{flalign}
&\mathit{A\_metadata}
= \mathit{tileM}\times\mathit{scale\_bits}_{k}
= \sqrt{8\,\mathit{NT}}\times
\tfrac{32\,w\,\sqrt{8\,\mathit{NT}}}{d\,k}
= \tfrac{256\,w\,\mathit{NT}}{d\,k} &\\
&\mathit{Since,}\;\mathit{tileN}
= \left\{\mathit{tileM}, \tfrac{tileM}{2}
\right\},\quad
\mathit{B\_metadata}
\in \left\{
\tfrac{256\,w\,\mathit{NT}}{d\,k},
\tfrac{128\,w\,\mathit{NT}}{d\,k}
\right\} &\\
&\mathit{total\_metadata}
= \mathit{A\_metadata}+\mathit{B\_metadata}
\in \left\{
\tfrac{512\,w\,\mathit{NT}}{d\,k},
\tfrac{384\,w\,\mathit{NT}}{d\,k}
\right\} &\\
&\mathit{per\_thread\_metadata}
= \tfrac{\mathit{total\_metadata}}{\mathit{NT}}
\in \left\{
\tfrac{512\,w}{d\,k},
\tfrac{384\,w}{d\,k}
\right\}
\mathit{bits} &
\end{flalign}

\hrule
\caption{Required Per-Thread MX Metadata Derivation}
\label{fig:mx_metadata_derivation}
\end{figure}

\paragraph{\textbf{MX Metadata Delivery and Storage}}
Since Vortex TCU WMMA instructions only expose three source operands, MX metadata cannot be supplied as an additional explicit source. RoadBlock instead duplicates registers \texttt{f8,f9} into a lightweight TCU-SRAM through a MUX activated when the WMMA's format field indicates an MX format. The scoreboard treats these as ``hidden source registers,'' to preserve dependency tracking. Additional metadata required by unorthodox format proposals can also be stored in this TCU-SRAM.

\begin{comment}
This contrasts with NVIDIA Blackwell, where private Tensor Memory (TMEM) stores both accumulators and scale metadata for \texttt{tcgen05.mma}, restricting maximum MMA tile utilization without complex warp-specialized pipelining \cite{Sul2025, VegaMyhre2026}. 
AMD CDNA4 instead stores MX scale matrices in dedicated VGPRs, increasing register pressure. RoadBlock provides a third design point by keeping MX metadata in TCU-SRAM while retaining only accumulators in the register file.
\end{comment}

\section{RoadBlock Host-Kernel ABI}
\begin{figure}[h]
    \centering
    \includegraphics[width=\linewidth]{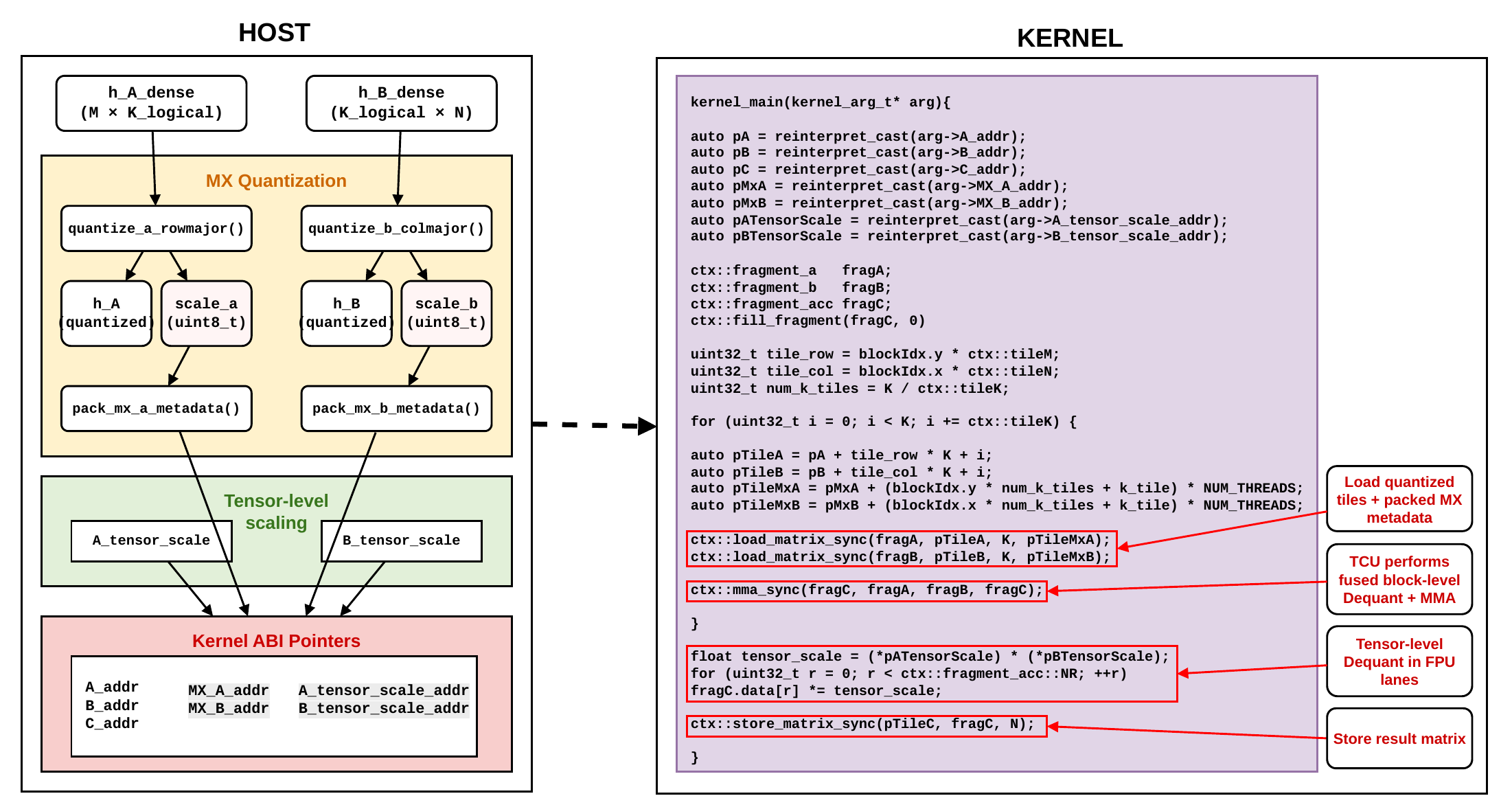}
    \caption{RoadBlock Host-Kernel ABI Flow}
    \label{fig:roadblock_hkabi}
\end{figure}

RoadBlock extends the Vortex WMMA ABI with pointers to the A/B scale-factor buffers produced during host-side quantization, as shown in Fig.~\ref{fig:roadblock_hkabi}. %we also prototyped interleaving scale metadata with the corresponding data tiles, but additional indexing/decode overhead largely negated the reduced kernel-interface overhead. 
For two-level formats such as NVFP4, block-wise E4M3 scaling is handled within the TCU, while tensor-wise FP32 dequantization is executed as WMMA epilogue on the FPU lanes.

\section{Experiments \& Evaluation}

To evaluate the framework's effectiveness, we modeled MXFP8, MXFP4, NVFP4, 4over6, IF4, and RaZeR on RoadBlock, and synthesized format-specific FEDP units using ASAP7 LVT/TT in Yosys under a 1.5\,GHz timing constraint. Fig.~\ref{fig:roadblock-area-power-tradeoff} characterizes Llama3.1-8B and Qwen3-32B perplexity against the formats' hardware efficiency across standard GPU Tensor Core scaling points ($N \in \{2,4,8,16\}$).

We find that, while MXFP4 has the worst perplexity among the group, its simple single-level E8M0 scale-factor and 32-element block size still make it the most efficient datatype when accounting for both accuracy and hardware cost together. On the other hand, while IF4 preserves model quality quite well, its extra range-alignment correction multiplier makes it an impractical addition to existing Tensor Core FEDP datapaths. As a middle-ground, RaZeR delivers the strongest Perplexity-PPA tradeoff among its NVFP4 counterparts, becoming especially attractive at larger FEDP lengths, exactly where modern GPU Tensor Core designs are headed.

\section{Future Work \& Conclusion}
Next, we plan on prototyping more unconventional emerging MX formats on RoadBlock and investigating warp-specialized kernels for multi-level block-scaled formats. In conclusion, RoadBlock provides a flexible, open-source, hardware-grounded means to evaluate whether a custom format proposal's algorithmic advantage survives integration into a realistic, programmable GPU Tensor Core setting.

\bibliographystyle{ACM-Reference-Format}
\bibliography{references}

\end{document}